\documentclass[aps,prl,twocolumn,showpacs,superscriptaddress,groupedaddress]{revtex4-2}  
\usepackage{graphicx}  
\usepackage{dcolumn}   
\usepackage{bm}        
\usepackage{amssymb}   
\usepackage{amsmath}   

\usepackage{verbatim}  

\usepackage{color}
\usepackage{soul}
\usepackage{comment}

\begin{document}
\title{Towards a theory of types III and IV non-Hermitian Weyl fermions}

\author{Zaur Z. Alisultanov}
\affiliation{Institute of Physics of DFRS, Russian Academy of Sciences, Makhachkala, Russia}

\author{Edvin G. Idrisov}
\affiliation{Department of Physics and Materials Science, University of Luxembourg, Luxembourg}
\date{\today}




\begin{abstract}
We develop the non-Hermitian Hamiltonian formalism to describe Weyl fermions of type III and IV. The spectrum of Hamiltonian has an unusual type of anisotropy. Namely, the hermiticity of Hamiltonian strongly depends on the direction in momentum space: for some directions the spectrum is real, in contrast for other directions it becomes complex. This fact leads to non-trivial adiabatic evolution and fractional Chern number. Additionally, we demonstrate that the non-Hermitian Hamiltonian can be regarded as a one-particle problem in context of topological band theory. 
\end{abstract}

\pacs{}
\maketitle


The current theoretical progress in band theory allows to describe topologically protected quasiparticles in modern experimentally accessible materials~\cite{Bansil}. Due to their unique properties, for instance chiral transport, the topological materials are expected to be promising for future electronics~\cite{Gilbert}. Among all, the topological materials with Dirac~\cite{Liu,Borisenko,Yang,Xu} and Weyl~\cite{Vafek,Weng,Huang,Nasser,Lv} points, as well as degenerate points~\cite{Bradlyn} and lines~\cite{Bzdusek,Rui,Bian} in Brillouin zone are of particular interest. The massless fermions in such materials are topologically protected, which results in quantum-electrodynamics effects known from high energy physics~\cite{VolovikBook}.

The idea of realization of Weyl points in condensed matter physics belongs back to C. Herring~\cite{Herring}. According to Herring's description, the presence of Weyl points can be understood within two-band model. The corresponding Hamiltonian of in momentum representation can be presented in Pauli basis
\begin{equation}
\label{Hamiltonain of two-band model in Pauli basis}
H(\mathbf{p})=\sigma_0 f_0(\mathbf{p})+\sum^3_{i=1}\sigma_i f_i(\mathbf{p}),
\end{equation}
where $\sigma_0$ is $2\times 2$ unit matrix, $\sigma_i$ are Pauli matrices and $\mathbf{p}$ is a three-dimensional (3D) momentum vector. The Weyl point $\mathbf{p}_{\text{W}}$ arises due to crossing of bands, which satisfies to three algebraic equations $f_i(\mathbf{p}_{\text{W}})=0$. Materials with these conditions are known as Dirac and Weyl semimetals~\cite{Armitage}. Typically such kind of situation is easily realizable in 3D systems, and equations $f_i(\mathbf{p}_{\text{W}})=0$ represent the two-dimensional (2D) surfaces in momentum space: three closed 2D surfaces can have many crossing points. In 2D system this condition corresponds to the crossing of three lines. This is not trivial and in order to have the Weyl point in 2D system, the certain types of symmetries (for example, $C_{6\nu}$ point symmetry) are required~\cite{Armitage}. In the vicinity of Weyl point $\mathbf{p}_{W}$ one can expand the function $f_i(\mathbf{p})$ in Taylor series, which results in low-energy linear spectrum, namely $f_i(\mathbf{p}) \propto |\mathbf{p}|$, and corresponding excitations are known as massless (Weyl or Dirac) fermions~\cite{Armitage}. The Eq.~(\ref{Hamiltonain of two-band model in Pauli basis}) describes conventional (non-tilted) type I Weyl fermions for zero and constant values of $f_0(\mathbf{p})$.

Apart from Weyl semimetals of type I, it was proposed Weyl semimetals of type II~\cite{Zubkov,Soluyanov,Sun,Wang}. Nowadays such kind of materials are accessible experimentally and WTe$_2$ is one of the candidates for a realization of type II Weyl fermions~\cite{Belopolski}. In short the Hamiltonian for such systems can be modeled at the intersections of Fermi-pockets and the spectrum turns out to be tilted. The minimal Hamiltonian for this case has the following form
\begin{equation}
\label{Hamiltonian for Weyl semimetal II}
H(\mathbf{p})=v_F \boldsymbol{\sigma} \cdot \mathbf{p}+\sigma_0 \boldsymbol{\omega} \cdot \mathbf{p},
\end{equation}    
where $v_F$ is Fermi velocity and $\boldsymbol{\omega}$ is a tilt vector. For $v_F> \omega$ this Hamiltonian describes the Weyl semimetals of type I with tilted spectrum, and consequently for $v_F < \omega$ the Hamiltonian corresponds to type II Weyl semimetals, where $\omega=|\boldsymbol{\omega}|$ is a modulus of vector. It is worth mentioning that Weyl semimetals of type II can be used for "modeling" of black and white holes, event horizon~\cite{Kedem,Thomas}. Indeed, assuming that the parameter of tilting, $\boldsymbol{\omega}(\mathbf{x})$, is a function of spatial coordinates, the Weyl fermions of type II can be described by action of massless spinor field~\cite{Kedem}
\begin{equation}
\label{Minimal action for Weyl type II fermions}
\mathcal{S}_{II}=\int d^{4}x\left[i\overline{\psi}\left(\upsilon_{F}\gamma^{\mu}\partial_{\mu}+\gamma^{0}\boldsymbol{\omega}(\mathbf{x})\cdot\boldsymbol{\partial}\right)\psi\right],
\end{equation}     
where $\psi$ is a Dirac spinor and $\gamma^{\mu}$ is the Dirac matrix in the Weyl representation $\gamma^{\mu}=\left(\begin{array}{cc}
	0 & \sigma^{\mu}\\
	\tilde{\sigma}^{\mu} & 0
\end{array}\right)$ with $\sigma^{\mu}=\left(\sigma_{0},\bm{\sigma}\right)$ and $\tilde{\sigma}^{\mu}=\left(\sigma_{0},-\bm{\sigma}\right)$ and $\partial_{\mu}=(v^{-1}_F \partial_t, \boldsymbol{\partial})$.
This is the possible basic (minimalistic) action for the description of massless spinor field in curved space with metrics $ds^{2}=\left(|\boldsymbol{\omega}|^{2}-1\right)dt^{2}-2\boldsymbol{\omega}\cdot d\mathbf{x}dt+d\mathbf{x}\cdot d\mathbf{x}$, and allows to model phenomena in the vicinity of event horizon.

In this Letter, we make a further step to develop the theory for Weyl fermions of types III and IV. We show that these fermions can be described using the Hamiltonian formalism in the framework of non-Hermitian quantum theory. Moreover, the non-Hermitian Hamiltonian arises as a one-particle problem in the context of band theory of Weyl semimetals. In order to study the properties of this Hamiltonian and associated Hilbert space, the acceptable theory of non-Hermitian systems is presented below. The spectrum of Weyl Hamiltonian under consideration turns out to be anisotropic in momentum space, namely for some directions the spectrum is real and for other directions it is complex quantity. The necessary and sufficient conditions for the real spectrum of the general Hamiltonian under consideration are formulated. The main feature of system under consideration is that "left" and "right" eigenvalues of Hamiltonian coincide with each other, which results in non-trivial adiabatic evolution and fractional Chern number.
    
\textit{Weyl fermions of type III and IV.}  In recent seminal papers, Ref.~\cite{Nissinen,Nissinen2}, authors have suggested Weyl fermions of type III and IV. Briefly, using the tetrad formalism, there was provided the general action of massless spinor field of form  
\begin{equation}
\label{Action for Weyl III and IV types fermions}
\mathcal{S}=\int d^{4}x\sqrt{-g}\left[i\overline{\psi}\gamma^{\nu}e_{\nu}^{\mu}\partial_{\mu}\psi\right],
\end{equation} 
where $g$ is the determinant of the metric tensor and $e_{\nu}^{\mu}$ is the tetrad tensor. One can see from Eq.~(\ref{Action for Weyl III and IV types fermions}), that Weyl fermions of type I and II are particular cases of the general tetrad action theory. Indeed, if one keeps only diagonal elements of tetrad tensor $e_{\nu}^{\mu}=v_{F}\delta_{\nu}^{\mu}$, then we get the action for Weyl femions of type I. If one set $e_{\nu}^{\mu}=v_{F}\delta_{\nu}^{\mu}+\delta_{\nu}^{0}\omega^{i}\delta_{i}^{\mu}$, then we get the action for type II Weyl fermions presented in Eq.~(\ref{Minimal action for Weyl type II fermions}). However, in more general case the Eq.~(\ref{Action for Weyl III and IV types fermions}) can contain other non-zero components of tetrad tensor, namely $e_{\nu}^{\mu}=v _{F}\delta_{\nu}^{\mu}+\delta_{\nu}^{0}\omega^{i}\delta_{i}^{\mu}+\vartheta_{i}\delta_{\nu}^{i}\delta_{0}^{\mu}$, where $\vartheta_i$ are the components of additional tilt parameter vector, $\boldsymbol{\vartheta}$. The presence of last term restores the symmetry with respect to rearrangement of indices $\mu$ and $\nu$. Note that this form corresponds to a spatially isotropic tetrad when there is no $\delta_{\nu}^{i}\delta_{j}^{\mu}$ term. According to Ref.~\cite{Nissinen}, the case $v_F > \omega$ and $v_F < \vartheta$ corresponds to Weyl fermions of type III, and $v_F < \omega$ and $v_F < \vartheta$ is associated with type IV, where  $\vartheta=|\boldsymbol{\vartheta}|$. Due to the last term, $\vartheta_{i}\delta_{\nu}^{i}\delta_{0}^{\mu}$, the Lagrangian contains a new term with zero component of momentum $\sigma^{i}\vartheta_{i}p_{0}$ and $p_{\mu}=(p_0, p_i)$ with $p_0=\varepsilon/v_F$. Particularly, the possible origin of this term and consequently Weyl points of type III and IV were associated with many-particle effects and thus can be worked out from self-energy~\cite{Nissinen}. It is worth mentioning, that in framework of Hermitian quantum mechanics the Hamilton formalism does not allow to take into account this term.

Nevertheless, we insist that the Hamiltonian formalism can be constructed, but in the framework of non-Hermitian quantum mechanics. Moreover, the non-Hermitian Hamiltonian formalism allows the alternative explanation, which does not necessarily include many-particle effects as the possible mechanism of explanation. In order to introduce the Hamiltonian, we first construct the wave equation 
\begin{equation}
\label{Wave equation}
G^{-1}(p_0,p_i)\left|\Psi\right\rangle=0,
\end{equation} 
where for the spatially isotropic case, which contains all important physics, the Greens function associated with action, Eq.~(\ref{Action for Weyl III and IV types fermions}) is given by 
\begin{equation}
\label{Green function for type III and IV Weyl fermions}
G^{-1}\left(p_{0},p_{i}\right)=v_{F}\sigma^{0}p_{0}-v_{F}\sigma^{i}p_{i}-\omega^{i}p_{i}-\sigma^{i}\vartheta_{i}p_{0}.
\end{equation} 
Using the above expression, the Eq.~(\ref{Wave equation}) can be easily rewritten in conventional form, where the Hamiltonian at left hand side is the function of only three dimensional momentum. After simple algebra one arrives to the following two wave equations 
\begin{equation}
\label{Shrodinger equation for R and L states}
\mathcal{H}\left|\Psi^{R}\right\rangle =\varepsilon_{R}\left|\Psi^{R}\right\rangle, \quad \mathcal{H}^{\dagger}\left|\Psi^{L}\right\rangle =\varepsilon_{L}\left|\Psi^{L}\right\rangle,
\end{equation} 
where the non-Hermitian Hamiltonian takes form 
\begin{equation}
\label{Non-Hermitian Hamiltonain}
\begin{split}	
& \mathcal{H}=(i\Gamma+K)/(1-\beta^2), \quad \Gamma=\boldsymbol{\sigma}\cdot\mathbf{\left[\mathbf{p}\times\boldsymbol{\vartheta}\right]}, \\
& K=\left[v_{F}\boldsymbol{\sigma}+\sigma_{0}(\boldsymbol{\omega}+\boldsymbol{\vartheta})\right] \cdot \mathbf{p}+\left(\boldsymbol{\sigma}\cdot\boldsymbol{\vartheta}\right)\left(\boldsymbol{\omega}\cdot\mathbf{p}\right)/v_F,
\end{split}
\end{equation}
and $\beta=\vartheta/v_F$. This is the Hamiltonian of type III and IV non-Hermitian Weyl fermions and is our main result. This Hamiltonian can be also obtained directly from the action, Eq.~(\ref{Action for Weyl III and IV types fermions}), if one writes the wave equation as the Euler-Lagrange equation for the Dirac field. In Eq.~(\ref{Shrodinger equation for R and L states}) we have defined the $R$ (right) and $L$ (left) wave functions. It is worth mentioning, that use of so called dual basis $(R,L)$ is a convenient step for non-Hermitian systems, which allows to restore the common structure of Hilbert space~\cite{Brody,Emil,Ghatak}. 

Now one can easily obtain the spectrum of non-Hermitian and conjugated Hamiltonians from Eq.~(\ref{Shrodinger equation for R and L states}). The spectra are coincides with each other, namely  
\begin{equation}
\label{Spectrum. Main formula}
\begin{split}
& \varepsilon=\varepsilon_{R}=\varepsilon_{L}=\frac{\mathcal{A}(\mathbf{p})\pm\sqrt{\mathcal{B}(\mathbf{p})}}
{1-\beta^2}, \quad \mathcal{A}(\mathbf{p})=(\boldsymbol{\vartheta}+\boldsymbol{\omega})\cdot\mathbf{p},\\
& \mathcal{B}(\mathbf{p})=[(\boldsymbol{\vartheta}+
\boldsymbol{\omega})\cdot\mathbf{p}]^{2}+(1-\beta^2)\left[v_{F}^{2}p^{2}-(\boldsymbol{\omega}\cdot\mathbf{p})^{2}\right],
\end{split}
\end{equation}  
where $p=|\mathbf{p}|$. Note, that the Hamiltonian of form $\mathcal{H}=a\sigma_{0}+\left(\mathbf{b}+i\mathbf{c}\right)\boldsymbol{\sigma}$ has the spectrum $\varepsilon=a\pm\sqrt{\mathbf{b}^{2}-\mathbf{c}^{2}+2i\mathbf{b}\cdot\mathbf{c}}$, which contains exceptional point due to term $\mathbf{b} \cdot \mathbf{c}$. In our case $\mathbf{b}$ is proportional (collinear) to $\mathbf{p}$ and $\mathbf{c}$ is proportional to vector product $[\mathbf{p} \times \boldsymbol{\vartheta}]$, thus $\mathbf{b} \cdot \mathbf{c}=0$ and the expression under the root always is real.
\begin{figure}
\includegraphics[width=\columnwidth]{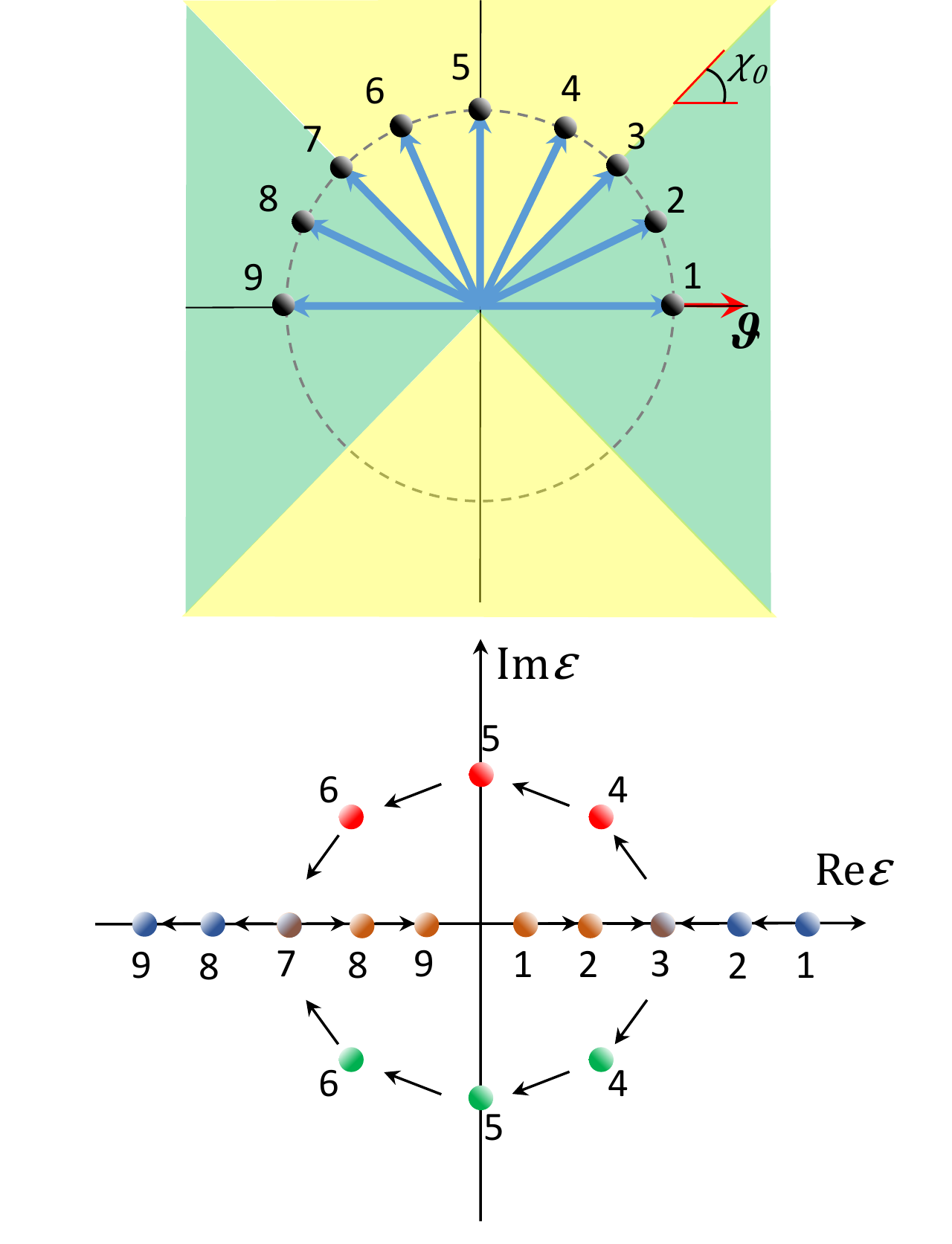}
\caption{\label{fig:two}
\textit{Top:} It is shown the segments of angles between vectors $\bm{\vartheta}$ (red arrow) and $\mathbf{p}$ (blue arrow), where the imaginary part of spectrum, $\textrm{Im}(\varepsilon)$ is zero (green) and non-zero (yellow). The angle $\chi_{0}=\arcsin(1/\beta)$ corresponds to the direction of the momentum vector relative to vector $\bm{\vartheta}$, at which the transition from the real to complex spectrum occurs. The black dots represent different directions of the momentum relative to vector $\bm{\vartheta}$. \textit{Bottom:} It is shown the evolution (black arrows) of the real and imaginary parts of the spectrum Eq.~(\ref{Spectrum. Main formula}) when the momentum direction changes from position $1$ to position $9$ (see the upper part of the figure). The brown and blue dots correspond to the hole and electronic states, while the red and green dots correspond to the complex spectrum with positive and negative imaginary parts. The dots $3$ and $7$ correspond to the transition points from the real to complex spectrum in Eq.~(\ref{Spectrum. Main formula}) at vanishing $\mathcal{B}\left(\mathbf{p}\right)$.}	
\end{figure}
For simplicity, we set $\boldsymbol{\omega}=0$ for further discussion. This does not affect on outcomes, since the non-hermiticity is described by term $i\Gamma$. Then Hamiltonian from Eq.~(\ref{Non-Hermitian Hamiltonain}) can be rewritten in the form 
\begin{equation}
\label{Hamiltonian in new form with unit vectors}
\mathcal{H}=\left[\boldsymbol{\sigma}\cdot\left(v_{F}\mathbf{e}_{p}+i\vartheta\sin\chi\mathbf{e}_{\chi}\right)p+\boldsymbol{\vartheta}
\cdot\mathbf{p}\right]/(1-\beta^2),
\end{equation} 
where the unit vector $\mathbf{e}_{p}$ is directed along the momentum $\mathbf{p}$ and the unit vector $\mathbf{e}_{\chi}$ is directed along the direction of cross product $[\mathbf{p} \times \boldsymbol{\vartheta}]$. For values $\vartheta < v_F$ ($\beta <1$) the parameter $\vartheta$ renormalizes the velocity $\partial \varepsilon/\partial \boldsymbol{p}$ due to coefficient $1/(1-\beta^2)$ and  results in slope (tilt). For $\vartheta > v_F$ ($\beta>1$), the spectrum in Eq.~(\ref{Spectrum. Main formula}) is real in segment $\left|\sin\chi\right|<\beta^{-1}$. In contrast, for  $\left|\sin\chi\right|> \beta^{-1}$ the spectrum consists of complex-conjugated branches. Therefore, we get the anisotropy with respect to spectrum's (Hamiltonian's) hermiticity (see Fig.~\ref{fig:two}). The eigenstates $\left|\Psi^{R}\right\rangle$  and $\left|\Psi^{L}\right\rangle$ are orthogonal in the segment $\left|\sin\chi\right|>\beta^{-1}$, where $\chi$ is the angle between vectors $\boldsymbol{\vartheta}$ and $\mathbf{p}$, i.e. $\left\langle \Psi^{L}|\Psi^{R}\right\rangle=0$. Indeed, from Eqs.~(\ref{Shrodinger equation for R and L states}), one has $\left\langle \Psi^{L}\right|\mathcal{H}\left|\Psi^{R}\right\rangle =\varepsilon_{R}\left\langle \Psi^{L}|\Psi^{R}\right\rangle =\varepsilon_{L}^{\ast}\left\langle \Psi^{L}|\Psi^{R}\right\rangle$. Since $\varepsilon_{R}\neq\varepsilon_{L}^{\ast}$, then $\left\langle \Psi^{L}|\Psi^{R}\right\rangle =0$. In the opposite segment $\left|\sin\chi\right|<\beta^{-1}$, where spectrum is real, generally speaking, these states are not orthogonal. Additionally, in Weyl point these states are not orthogonal as well, since $\varepsilon_R=\varepsilon_L=0$. Further, it is easy to show, that in case of real spectrum $\left\langle \varPsi_{\pm}^{L}|\varPsi_{\mp}^{R}\right\rangle =0$ and $\left\langle \varPsi_{\pm}^{L}||\varPsi_{\pm}^{R}\right\rangle \neq0$, where "$\pm$" corresponds to electron hole states in Eq.~(\ref{Spectrum. Main formula}). At the same time, in the segment, where spectrum is complex one has  $\left\langle \varPsi_{\pm}^{L}||\varPsi_{\pm}^{R}\right\rangle =0$ and $\left\langle \varPsi_{\pm}^{L}|\varPsi_{\mp}^{R}\right\rangle \neq0$. Such a situation leads to non-trivial adiabatic evolution (see Sec.~A and B in Ref.~\cite{SM}). The existence of segment with real spectrum, despite the non-Hermitian nature of Hamiltonian, is surprising. Nevertheless, it can be proven the following statement:  \textit{If the non-Hermitian Hamiltonian, $\mathcal{H}$ can be brought to hermitian $\mathcal{H}_{h}$ one, using the similarity transformations with non-degenerate matrix, then the eigenvalues of $\mathcal{H}$ are real.} Indeed, let apply the similarity transformation, $S\mathcal{H}S^{-1}$ with the linear operator $S$ in the from of non-degenerate matrix, $(S^{-1})^{\dagger}=(S^{\dagger})^{-1}$, to Hamiltonian $\mathcal{H}$. Next, we assume that the following equality holds
\begin{equation}
\label{Similarity condition}
S\mathcal{H}S^{-1}=\mathcal{H}_{h}.
\end{equation} 
Then $S\mathcal{H}S^{-1}\left|\Psi\right\rangle =\mathcal{H}_{h}\left|\Psi\right\rangle =\mathcal{E}\left|\Psi\right\rangle$ , where $\mathcal{E}$ is real since the Hamiltoanin $\mathcal{H}_{h}$ is hermitian. On the other hand due to linearity of operator $S$, this equation can be written as $\mathcal{H}\left|\overline{\Psi}\right\rangle =\mathcal{E}\left|\overline{\Psi}\right\rangle$, where $\left|\overline{\Psi}\right\rangle =S^{-1}\left|\Psi\right\rangle$. Thus, the eigenvalues of non-Hermitian Hamiltonian, which satisfies the condition Eq.~(\ref{Similarity condition}), are real. The Eq.~(\ref{Similarity condition}) can be rewritten in new form 
\begin{equation}
\label{Similarity condition two}
\mathcal{H}^{\dagger}=\left(S^{\dagger}S \right)\mathcal{H}\left(S^{\dagger}S\right)^{-1}=\eta \mathcal{H}\eta^{-1},
\end{equation}
where we have introduced the Hermitian operator $\eta=S^{\dagger}S$. The Eq.~(\ref{Similarity condition two}) is the necessary and sufficient condition for the spectrum of Hamiltonain $\mathcal{H}$ to be real. If $S$ is unitary matrix, then $S^{\dagger}S=1$ and Eq.~(\ref{Similarity condition two}) recovers the hermicity of Hamiltonian from hermitian quantum mechanics. One can show that the Hamiltonian Eq.~(\ref{Hamiltonian in new form with unit vectors}) satisfies the condition (\ref{Similarity condition two}) for $|\sin \chi| < \beta^{-1}$ segment (see Sec.~C of Ref.~\cite{SM}). For non-degenerate matrix $\left(S^{-1}\right)^{\dagger}=\left(S^{\dagger}\right)^{-1}$, the condition in Eq.~(\ref{Similarity condition two}) coincides with the condition of pseudo-hermiticity~\cite{Ali1,Ali2,Ali3}. This is the necessary condition for spectrum to be real. The sufficient condition for spectrum to be real is the existence of operator $\eta$ in the form of $S^{\dagger}S$. The condition in Eq.~(\ref{Similarity condition two}) at $S^{\dagger}S=\eta$ is known as $\eta-$ hermiticity. We call the Hamiltonains $\mathcal{H}$ and $\mathcal{H}_h$ from Eq.~(\ref{Similarity condition}) equivalent, since they have the same real spectrum even if one of them is non-Hermitian. The equivalence of two Hamiltonians in that sense is the necessary and sufficient condition to have real valued spectrum. 

\textit{Topological protection.} We continue to investigate the spectrum of Weyl fermions of types III and IV. It is worth mentioning, that Weyl points are topologically protected with respect to external perturbations. The spectrum remains gapless, for the case of unit matrix perturbation, $H \to H+IU_{0}$, which causes the shift of Weyl points with respect to energy and momentum, $\varepsilon \to \varepsilon (p-p_{0})+ \varepsilon (p_{0})$, where $p_0$ satisfies to the condition $ \left[\left(\boldsymbol{\vartheta}+\boldsymbol{\omega}\right)\cdot\mathbf{p_{0}}+U_{0}\right]^{2}+\left[1-\beta^2\right]\left[v_{F}^{2}p_{0}^{2}-\left(\boldsymbol{\omega}\cdot\mathbf{p_{0}}+U_{0}\right)^{2}\right]=0$. Next, the perturbation in the form of Pauli matrix, $H \to H+\boldsymbol{\sigma}\mathbf{U}$ results in the following spectrum 
\begin{equation}
\label{Topological protection.Spectrum with Pauli matrix perturbation}
\varepsilon=\left[\mathcal{E}_1(\mathbf{p})\pm \mathcal{E}_2(\mathbf{p})\right]/\left(1-\beta^2 \right), \\	
\end{equation}
where $\mathcal{E}_1(\mathbf{p})=\mathcal{A}(\mathbf{p})+\boldsymbol{\vartheta} \cdot \mathbf{U}/v_F$ and $\mathcal{E}_2(\mathbf{p})=\sqrt{\mathcal{E}^2_1(\mathbf{p})+(1-\beta^2)\xi(\mathbf{p})}$, with $\xi(\mathbf{p})=(v_{F}\mathbf{p}+\mathbf{U})^{2}-\left(\boldsymbol{\omega}\cdot\mathbf{p}\right)^{2}$ and $\mathcal{A}(\mathbf{p})$ is given in Eq.~(\ref{Spectrum. Main formula}). This spectrum has a shift with respect to initial one, Eq.~(\ref{Spectrum. Main formula}), but does not contain a gap.
\begin{figure}
\includegraphics[width=\columnwidth]{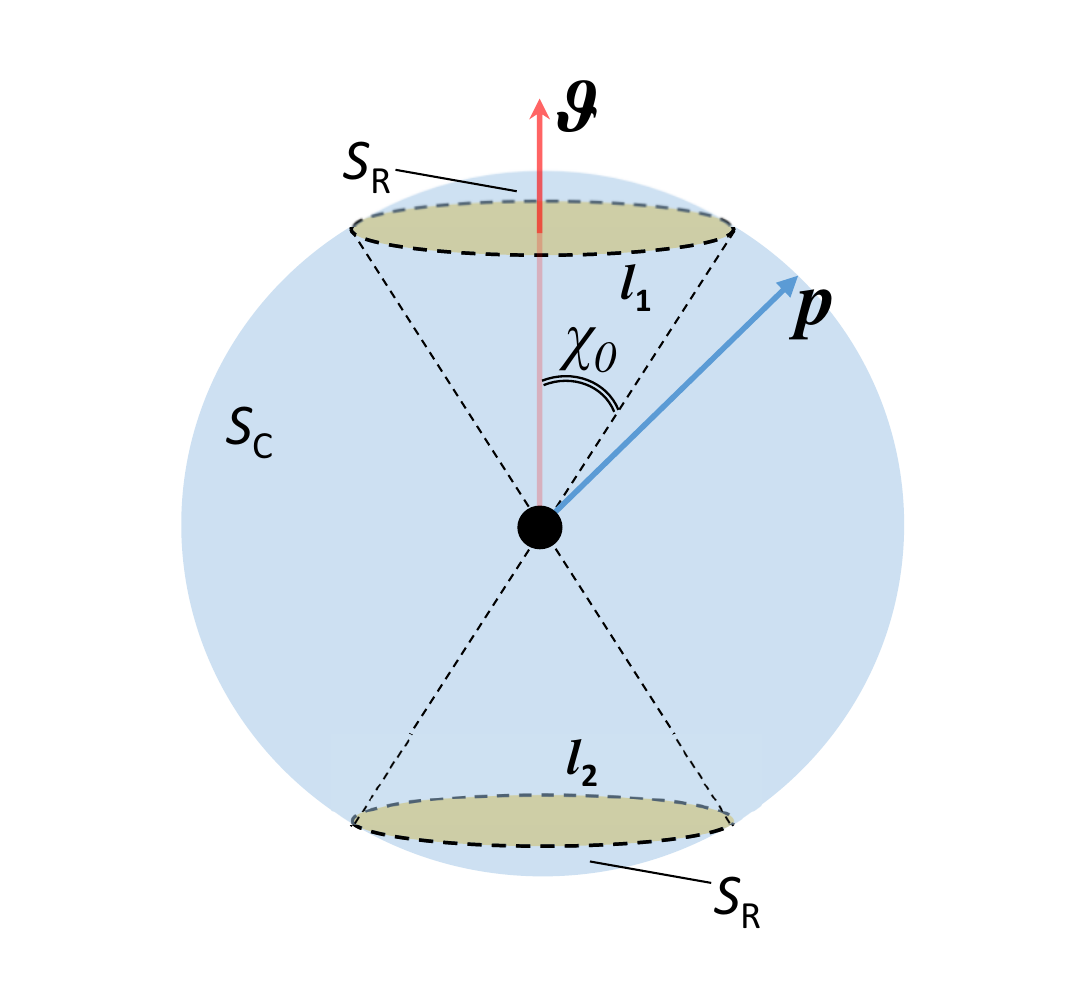}
\caption{\label{fig:four} It is shown a closed surface in momentum space enclosing the Weyl point. The surface is divided into two regions: The spectrum of Hamiltonian, Eq.~(\ref{Hamiltonian in new form with unit vectors}) is real in the region $S_{\textrm {R}}$ and complex in the region $S_{\textrm {C}}$. The lines $\bm{l}_{1}$ and $\bm{l}_{2}$ represent the boundaries between $S_{\textrm {R}}$ and $S_{\textrm {C}}$. The vector $\bm{\vartheta}$ is directed along $Z$-axis.
}	
\end{figure}

It is as well  useful to investigate the stability of spectrum with the help of topological invariant-Chern number. For non-Hermitian Hamiltonian the Chern number can be written as~\cite{Fu,Ghatak}
\begin{equation}
\label{Chern number}
\mathcal{N}=\frac{1}{4\pi}\oint i\partial_{\mathbf{p}}\times\left\langle \varPsi^{L}\right|\partial_{\mathbf{p}}\left|\varPsi^{R}\right\rangle \cdot d\mathbf{S},	
\end{equation}
where the integration is carried out over a closed surface enclosing the Weyl point. The Chern number defined in Eq.~(\ref{Chern number}) must vanish, if the Hamiltonian has certain symmetries, namely $\mathcal{H}\left(\mathbf{p}\right)=\mathcal{H}^{T}\left(\mathbf{p}\right)$, $\sigma_{x}\mathcal{H}\left(\mathbf{p}\right)\sigma_{x}=\mathcal{H}^{T}\left(\mathbf{p}\right)$ or $\mathcal{H}\left(\mathbf{p}\right)=\mathcal{H}^{T}\left(-\mathbf{p}\right)$. The Hamiltonian does not satisfies any of these conditions, which indicates of topological protection of Chern invariant for the Hamiltonian under consideration.

The integration domain in Eq.~(\ref{Chern number}) can be divided into two regions, $S_{\textrm{C}}$ and $S_{\textrm{R}}$, which correspond to complex and real spectrum of Hamiltonian in Eq.~(\ref{Hamiltonian in new form with unit vectors}) (see Fig.~\ref{fig:four}). The region $S_{\textrm{C}}$ does not contribute to integral in Eq.~(\ref{Chern number})~(see Sec.D of Ref.~\cite{SM}), therefore $\mathcal{N}=\frac{1}{4\pi}\int_{S_{\textrm{R}}}i\partial_{\mathbf{p}}\times\left\langle \varPsi^{L}\right|\partial_{\mathbf{p}}\left|\varPsi^{R}\right\rangle \cdot d^{2}\mathbf{p}$, which results in fractional Chern number. This means that Weyl fermions of types III and IV are characterized with fractional Chern number. Fractional topological indexes have been considered earlier in literature, and typically are associated with exceptional points in spectrum~\cite{TonyLee}, in contrast to the case under consideration with no exceptional points.

\begin{figure}	
	\includegraphics[width=\columnwidth]{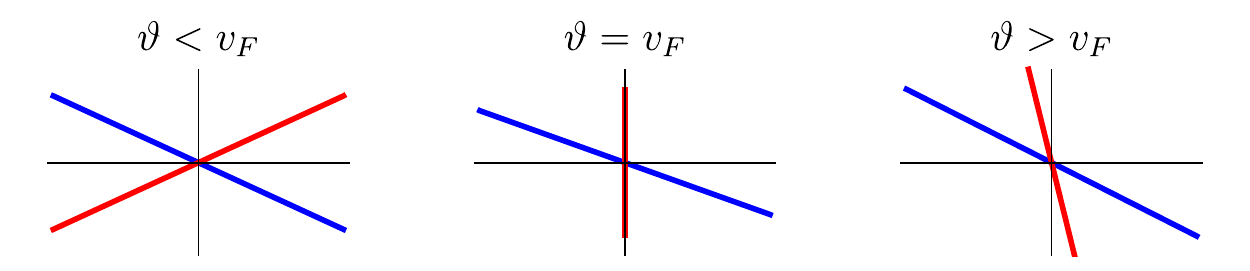}
	\caption{\label{fig:figure3} The electron spectrums of type I (left) and III (right). The vertical axis is energy and the horizontal axis is momentum. The spectrum during the phase transition I-III (middle).} 
\end{figure}

\textit{Discussion.} Let investigate the spectrum in phase transition point $v_F=\vartheta$. For simplicity, we assume that vectors $\boldsymbol{\vartheta},\boldsymbol{\omega}$ are parallel to $\mathbf{p}$. Then one has $\left(\boldsymbol{\vartheta}+\boldsymbol{\omega}\right)\cdot\mathbf{p}=\pm\left(\vartheta+\omega\right)p$, where "$\pm$" correspond to positive and negative directions of momentum $\mathbf{p}$. Let be $v_F> \omega$. This corresponds to phase transition between type I and III. The electron states with positive momentum and hole states with negative momentum in phase transition point can be associated with infinite group velocity $\partial\varepsilon/\partial p$.
Indeed, in this case from Eq.~(\ref{Spectrum. Main formula}) one has 
\begin{equation}
	\label{Infinite Fermi velocity}
	\lim_{\vartheta \to v_{F}}\varepsilon\rightarrow\infty.
\end{equation} 
Such situation is described by vertical line. The second branch of spectrum originates from electron states with negative momentum and hole states with positive momentum. For these states from Eq.~(\ref{Spectrum. Main formula}) we have
\begin{equation}
	\label{Finite Fermi velocity}
	\lim_{\vartheta \to v_F} \varepsilon=-\frac{v^2_F p^2-(\boldsymbol{\omega} \cdot \mathbf{p})^2}{2(\boldsymbol{\vartheta}+\boldsymbol{\omega}) \cdot \mathbf{p}}.
\end{equation}
These two branches for the cases $v_F> \vartheta$, $v_F=\vartheta$ and $v_F< \vartheta$ are provided at Fig.~\ref{fig:figure3} in case of type III (the qualitative picture for the case of type IV is similar).
Let discuss these transitions in details. To do this, let consider the linear function $y=(a-b)x$. For $a>b$ this function describes the line with positive slope. At point $a=b$ the continuous transition from positive to negative slope occurs. At the same time the point $a=b$ corresponds to horizontal line, i.e. the transition occurs through $X$ axis. Is it possible to make a phase transition from positive slope to negative through $Y$ axis? This transition correspond to type I-III (II-IV) and it is completely different compared to previous one. The main difference is that during the phase transition there is the swapping of the electron and hole states. Such kind of transition can be described with function of form $y=x/(a-b)$. For $a>b$ this function describes the straight line with positive slope. For $a<b$ the slope is negative. The point $a=b$ correspond to transition from positive to negative slope. However, compared to previous example, $(a-b)x$, this is the singularity point. Thus, the transition from positive to negative slope is not continuous.  In other words, in second example conditions $a>b$ and $a<b$ correspond to fundamentally different phases, and the continuous transition between them is forbidden. Now, it become clear that types I and II are not topologically different since the continuous transition is allowed. Consequently, in this context, types I and III, I and IV are topologically different phases. Note that the discussion above is true only for one branch (in our case it is red line). For blue branch there is no inversion of electron and hole parts of spectrum. Indeed, the slope of blue branch of spectrum does not depend on the ratio between $v_F$ and $\vartheta$.       

Finally, let us discuss the possible origin of non-Hermitian Weyl Hamiltonian of types III and IV. The authors of Ref.~\cite{Nissinen} suggest to associate the Weyl fermions of types III and IV with many-particle effects. However, it can be provided the alternative  interpretation as well. The alternative interpretation does not depend directly on the many-particle effects. We present the explanation in a general form, keeping in mind two-band model of band structure. The wave function in two-band model has the form $\left|\Psi_{\mathbf{p}}\right\rangle =C_{u}\left|u_{\mathbf{p}}\right\rangle +C_{v}\left|v_{\mathbf{p}}\right\rangle$, where amplitude $C_{u/v}$ corresponds to $\left|u_{\mathbf{p}}\right\rangle / \left|u_{\mathbf{p}}\right\rangle$ Bloch's function.
Multiplying the wave equation $	H\left|\Psi_{\mathbf{p}}\right\rangle=E_{\mathbf{p}}\left|\Psi_{\mathbf{p}}\right\rangle$ with $\left\langle u_{\mathbf{p}}\right|$ and $\left\langle \upsilon_{\mathbf{p}}\right|$, we obtain the system of equations for the amplitudes 
\begin{equation}
\label{System of equations for two-band model}
\begin{split}
&	
\left(\begin{array}{cc}
H_{uu}^{\mathbf{p}} & H_{uv}^{\mathbf{p}}\\
H_{vu}^{\mathbf{p}} & H_{vv}^{\mathbf{p}}
\end{array}\right)\left(\begin{array}{c}
C_{u}\\
C_{v}
\end{array}\right)=E_{\mathbf{p}}\left(\begin{array}{cc}
		S_{uu}^{\mathbf{p}} & S_{uv}^{\mathbf{p}}\\
		S_{v u}^{\mathbf{p}} & S_{vv}^{\mathbf{p}}
	\end{array}\right)\left(\begin{array}{c}
		C_{u}\\
		C_{v}
	\end{array}\right) \\
& =\left[f_{0}(\mathbf{p})+\sigma_{i}f_{i}(\mathbf{p})\right]\left(\begin{array}{cc}
	C_u & C_v\\
\end{array}\right)^T,
\end{split}
\end{equation} 
where $H_{ab}^{\mathbf{p}}=\left\langle a_{\mathbf{p}}\right|H\left|b_{\mathbf{p}}\right\rangle $ is the matrix element of Hamiltonian, with $a,b=u,v$ and $S_{ab}^{\mathbf{p}}=\left\langle a_{\mathbf{p}}|b_{\mathbf{p}}\right\rangle =\int a_{\mathbf{p}}^{\ast}\left(\mathbf{r}\right)b_{\mathbf{p}}\left(\mathbf{r}\right)d^{3}\mathbf{r}$ is the overlap integral. Typically, the Bloch functions are taken to be orthogonal, thus in Eq.~(\ref{System of equations for two-band model}) the overlap matrix becomes unit matrix, $S^{\mathbf{p}}_{ab}=\delta_{ab}$. However, in real materials, this is not always true and off-diagonal elements of overlap matrix do not vanish (for instance, one of the reason is indeed many-particle effects)~\cite{Landsberg,Halpern,Bernstein,Boykin}. Let introduce the Bloch functions $u_{\mathbf{p}}(\mathbf{r})=\sum_{\mathbf{R}}e^{i\mathbf{p}\mathbf{R}}\phi_u(\mathbf{r}-\mathbf{R})$ and $v_{\mathbf{p}}(\mathbf{r})=\sum_{\mathbf{R}}e^{i\mathbf{p}\mathbf{R}}\phi_v(\mathbf{r}-\mathbf{R})$, where $\phi_u$, $\phi_v$ atomic orbitals, corresponding to bands $u$, $v$ and $\mathbf{R}$ is a radius vector of a given atom at lattice site. Then the off-diagonal overlap integral can be written as $\int u^{\ast}_{\mathbf{p}}(\mathbf{r}) v_{\mathbf{p}}(\mathbf{r}) d^3 \mathbf{r}=\sum_{\mathbf{R} \mathbf{R}^{\prime}} e^{i\mathbf{p}(\mathbf{R}-\mathbf{R}^{\prime})} \int \phi^{\ast}_u(\mathbf{r}-\mathbf{R}^{\prime}) \phi_v(\mathbf{r}-\mathbf{R})d^3 \mathbf{r}$. Atomic orbitals corresponding to the same atom ($\mathbf{R}=\mathbf{R}^{\prime}$) are orthogonal, $\int \phi^{\ast}_a(\mathbf{r}-\mathbf{R}^{\prime}) \phi_b(\mathbf{r}-\mathbf{R})d^3 \mathbf{r}=\delta_{ab}$. However, at different $\mathbf{R} \neq \mathbf{R}^{\prime}$, generally speaking these orbitals are not orthogonal, which results in non-zero off-diagonal elements of overlap matrix, $S^{\mathbf{p}}_{ab}$. At the same time, we note, that the Bloch functions at the Weyl point will be orthogonal, since at this point the time reversal symmetry is not broken. This means that $\left\langle u_{\mathbf{p}_{W}}|\mathcal{T}u_{\mathbf{p}_{W}}\right\rangle =0$, where $\mathcal{T}$ is the time reversal operator~\cite{Herring}. Since $\mathcal{T}\left|u_{\mathbf{p}_{W}}\right\rangle =\left|\upsilon_{\mathbf{p}_{W}}\right\rangle$  then $\left\langle u_{\mathbf{p}_{W}}|\upsilon_{\mathbf{p}_{W}}\right\rangle =0$. Further, the Eq.~(\ref{System of equations for two-band model}) can be rewritten as $(S^{\mathbf{p}})^{-1}\left[f_{0}(\mathbf{p})+\sigma_{i}f_{i}(\mathbf{p})\right]\left(\begin{array}{cc}
	C_u & C_v\\
\end{array}\right)^T=\tilde{\mathcal{H}}\left(\begin{array}{cc}
C_u & C_v\\
\end{array}\right)^T=\tilde{\varepsilon} \left(\begin{array}{cc}
C_u & C_v\\
\end{array}\right)^T$, where $S^{\mathbf{p}}_{ab}$ is a matrix on the right hand side of Eq.~(\ref{System of equations for two-band model}). In the vicinity of Weyl point one has $f_i(\mathbf{p}) \approx v_F p_i$ and $f_0(\mathbf{p}) \approx \boldsymbol{\omega} \cdot \mathbf{p}$. If one requires the overlap matrix in Pauli basis has components $S_{uu(vv)}^{\mathbf{p}}=1\mp \vartheta_{z}/v_F$, $S_{uv}^{\mathbf{p}}=(i\vartheta_{y}-\vartheta_{x})/v_F=(S_{v u}^{\mathbf{p}})^{\ast}$, then the effective Hamiltonian, $\tilde{\mathcal{H}}=(S^{\mathbf{p}}_{ab})^{-1}\left[f_{0}(\mathbf{p})+\sigma_{i}f_{i}(\mathbf{p})\right]$ completely coincides with the Hamiltonian from Eq.~(\ref{Non-Hermitian Hamiltonain}) with the same prefactor $1/(1-\beta^2)$. Thus the origin of this Hamiltonian can be related with the overlap between Bloch functions. For 1D chain such a Hamiltonian has been recently introduced by generalizing the Su-Schrieffer-Heeger model on non-hermitian case~\cite{Benjamin}.

To summarize, we have provided the theory of type III and IV Weyl semimetals within non-Hermitian Hamiltonian formalism. The spectrum of this Weyl Hamiltonian exhibits an unusual type of anisotropy in momentum space, namely for some directions the spectrum is real, in contrast for other directions it is complex. The necessary and sufficient conditions for spectrum to be real is provided. Moreover, the anisotropy in spectrum results in anisotropy of the inner product of states, and thus leads to non-trivial adiabatic evolution and fractional Chern number. Additionally, we speculate the possible origin of non-Hermitian Hamiltonian.     

We are grateful to G. E. Volovik, J. Nissinen and T. L. Schmidt for fruitful discussions. Z. A. acknowledges the financial support from President Grant (MD 647.2020.2). E. I. acknowledges financial support from the National Research Fund Luxembourg under Grant CORE C19/MS/13579612/HYBMES.


\setcounter{equation}{0}
\setcounter{figure}{0}
\renewcommand{\theequation}{S-\arabic{equation}}
\renewcommand{\thefigure}{S-\arabic{figure}}
\onecolumngrid
\newpage
\begin{center}
	{\large{\bf Supplementary Material for ``Towards a theory of types III and IV non-Hermitian Weyl fermions''}}\\
	\vspace{4mm}
	
\end{center}

\section{A: Anisotropy of orthogonality conditions}
\label{Sec:A}
We write the spectrum from Eq.~(9) in main text with additional subscripts as
\begin{equation}
	\varepsilon_{R,\pm}=\varepsilon_{L,\pm}=\frac{\mathcal{A}\left(\mathbf{p}\right)\pm\sqrt{\mathcal{B}\left(\mathbf{p}\right)}}{1-\beta^{2}},
\end{equation}
where "$\pm$" are associated with electron and hole states. The corresponding eigenvectors are  denoted by $\left|\varPsi_{\pm}^{R,L}\right\rangle$, which satisfy wave equations $ \mathcal{H}\left|\varPsi_{\pm}^{R}\right\rangle =\varepsilon_{R,\pm}\left|\varPsi_{\pm}^{R}\right\rangle$, and $
\mathcal{H}^{\dagger}\left|\varPsi_{\pm}^{L}\right\rangle =\varepsilon_{L,\pm}\left|\varPsi_{\pm}^{L}\right\rangle$. It is worth mentioning that $\varepsilon_{R}=\varepsilon_{L}=\varepsilon$, where "$\pm$" is omitted.
In phases III and IV the spectrum is anysotropic (see Eq.~(9) from main text) with respect to space orientation of momentum vector: for some directions the spectrum is real, in contrast, for other directions it is complex. The existence of domain, where $\text{Im}(\varepsilon)\neq 0$ is the main distinguishing feature of the new phases, III and IV. Apart from this, in domain of complex spectrum $\varepsilon_{\pm}^{*}=\varepsilon_{\mp}$, which leads to important consequences. The main one is related with scalar product $\left\langle \varPsi_{\alpha}^{L}|\varPsi_{\beta}^{R}\right\rangle$. Let consider the following $2 \times 2$ matrix
\begin{equation}
	\left(\begin{array}{cc}
		\left\langle \varPsi_{+}^{L}|\mathcal{H}|\varPsi_{+}^{R}\right\rangle  & \left\langle \varPsi_{+}^{L}|\mathcal{H}|\varPsi_{-}^{R}\right\rangle \\
		\left\langle \varPsi_{-}^{L}|\mathcal{H}|\varPsi_{+}^{R}\right\rangle  & \left\langle \varPsi_{-}^{L}|\mathcal{H}|\varPsi_{-}^{R}\right\rangle
	\end{array}\right).
\end{equation}
Applying the Hamiltonian inside matrix elements on right and left hand sides one obtains 
\begin{equation}
	\left(\begin{array}{cc}
		\varepsilon_{+}\left\langle \varPsi_{+}^{L}||\varPsi_{+}^{R}\right\rangle  & \varepsilon_{-}\left\langle \varPsi_{+}^{L}||\varPsi_{-}^{R}\right\rangle \\
		\varepsilon_{+}\left\langle \varPsi_{-}^{L}||\varPsi_{+}^{R}\right\rangle  & \varepsilon_{-}\left\langle \varPsi_{-}^{L}||\varPsi_{-}^{R}\right\rangle
	\end{array}\right)=\left(\begin{array}{cc}
		\varepsilon_{+}^{*}\left\langle \varPsi_{+}^{L}||\varPsi_{+}^{R}\right\rangle  & \varepsilon_{+}^{*}\left\langle \varPsi_{+}^{L}||\varPsi_{-}^{R}\right\rangle \\
		\varepsilon_{-}^{*}\left\langle \varPsi_{-}^{L}||\varPsi_{+}^{R}\right\rangle  & \varepsilon_{-}^{*}\left\langle \varPsi_{-}^{L}||\varPsi_{-}^{R}\right\rangle
	\end{array}\right).
\end{equation}
For real spectrum from the above equation it is obvious that $\left\langle \varPsi_{\pm}^{L}|\varPsi_{\mp}^{R}\right\rangle =0$ and $\left\langle \varPsi_{\pm}^{L}||\varPsi_{\pm}^{R}\right\rangle \neq 0$. At the same time, for the domain, where the spectrum is complex one gets $\left\langle \varPsi_{\pm}^{L}||\varPsi_{\pm}^{R}\right\rangle=0$ and $\left\langle \varPsi_{\pm}^{L}|\varPsi_{\mp}^{R}\right\rangle \neq 0$.

\section{B: The form of eigenstates}
The Hamiltonian under consideration has the form 
\begin{equation}
	\mathcal{H}=A\left(\mathbf{p}\right)\sigma_{0}+\left[\bm{D}\left(\mathbf{p}\right)+i\bm{C}\left(\mathbf{p}\right)\right]\cdot\bm{\sigma},
\end{equation}
where $\bm{D}\cdot\bm{C}=0$. The spectrum of this Hamiltonian is given by 
\begin{equation}
	\varepsilon_{R,\pm}=\varepsilon_{L,\pm}=\varepsilon_{\pm}=A\pm\sqrt{D^{2}-C^{2}},
\end{equation}
where $C=\left|\bm{C}\right|$ and $D=\left|\bm{D}\right|$. Particularly, for the case under consideration in main text, the parameters are $A=(\bm{\omega}\cdot\mathbf{p}+\bm{\vartheta}\cdot\mathbf{p})/(1-\beta^{2})$, $\bm{D}=(\upsilon_{F}\mathbf{p}+\frac{1}{\upsilon_{F}}\bm{\vartheta}\left(\bm{\omega}\cdot\mathbf{p}\right))/(1-\beta^{2})$ and  $\bm{C}=\left[\mathbf{p}\times\bm{\vartheta}\right]/(1-\beta^{2})$.
The eigenstates of Hamiltonian are given by 
\begin{equation}
	\left|\varPsi_{\pm}^{R}\right\rangle =\left(\begin{array}{c}
		\frac{C_{z}-iD_{z}\mp i\sqrt{D^{2}-C^{2}}}{C_{x}+i\left(C_{y}-D_{x}\right)+D_{y}}\\
		1
	\end{array}\right),\,\,\,\,\left|\varPsi_{\pm}^{L}\right\rangle =\left(\begin{array}{c}
		\frac{C_{z}+iD_{z}\pm i\sqrt{D^{2}-C^{2}}}{C_{x}+i\left(C_{y}+D_{x}\right)-D_{y}}\\
		1
	\end{array}\right),
\end{equation}
and
\begin{equation}
	\left\langle \varPsi_{\pm}^{R}\right|=\left(\begin{array}{cc}
		\frac{C_{z}+iD_{z}\pm i(\sqrt{D^{2}-C^{2}})^{*}}{C_{x}-i\left(C_{y}-D_{x}\right)+D_{y}} & 1\end{array}\right),\,\,\,\,\left\langle \varPsi_{\pm}^{L}\right|=\left(\begin{array}{cc}
		\frac{C_{z}-iD_{z}\mp i(\sqrt{D^{2}-C^{2}})^{*}}{C_{x}-i\left(C_{y}+D_{x}\right)-D_{y}} & 1\end{array}\right), 
\end{equation}
where "$\ast$" means the complex conjugation. Using the form of eigenstates one can show that $\left\langle \varPsi_{\pm}^{L}|\varPsi_{\mp}^{R}\right\rangle =0$ for $D>C$, and $\left\langle \varPsi_{\pm}^{L}|\varPsi_{\pm}^{R}\right\rangle =0$ for $D<C$. Further, the direct calculations give $\left\langle \varPsi_{\pm}^{L}|\varPsi_{\pm}^{R}\right\rangle =C^{2}-D^{2}-\left|D^{2}-C^{2}\right|\mp2i\left(C_{z}-iD_{z}\right)\textrm{Re}\sqrt{D^{2}-C^{2}}$. Therefore, for $D<C$ we get $\left\langle \varPsi_{\pm}^{L}|\varPsi_{\pm}^{R}\right\rangle =0$. On the other hand, $\left\langle \varPsi_{\mp}^{L}|\varPsi_{\pm}^{R}\right\rangle =C^{2}-D^{2}+\left|D^{2}-C^{2}\right|\pm2\left(C_{z}-iD_{z}\right)\textrm{Im}\sqrt{D^{2}-C^{2}}$. Thus, for $D>C$ we have $\left\langle \varPsi_{\mp}^{L}|\varPsi_{\pm}^{R}\right\rangle =0$.

\section{C: Real spectrum of pseudo-Hermitian Hamiltonian}
\label{Sec:C}
From the main text, Eq.~(12), it is clear, that $\left[\eta,\mathcal{H}\right]=\eta\mathcal{H}-\mathcal{H}\eta\neq 0$, for non-Hermitian $\mathcal{H}$. Indeed, otherwise $[\eta,\mathcal{H}]=0$ and one arrives to contradiction, namely the conjugated Hamiltonian $\mathcal{H}^{\dagger}=\eta\mathcal{H}\eta^{-1}=\mathcal{H}\eta\eta^{-1}=\mathcal{H}$ is Hermitian. As an example, let consider the simple non-Hermitian matrix $M=\left(\begin{array}{cc}
	1 & 1+\lambda\\
	1-\lambda & -1
\end{array}\right)$, where $\lambda$ is the arbitrary number. Next, let find the matrix $\eta$, which satisfies the condition $M^{\dagger}=\eta M\eta^{-1}$. One can show that the $\eta$ has the following form $\eta=\left(\begin{array}{cc}
	\frac{r+q}{1+\lambda} & q\\
	q & \frac{r-q}{1-\lambda}
\end{array}\right)$, where $p$, $q$ are arbitrary numbers as well. Further, in order to have the real eigenvalues for matrix $M$, there must be possibility to present $\eta=S^{\dagger}S$. From these condition, one consequently obtains that $q,r\in\mathbb{R}$ and $\eta^{\dagger}=\eta$. Further, it is obvious, that the condition $\det\left(SS^{\dagger}\right)>0$ must be satisfied. This condition set limits on the choice of elements of matrix $\eta$. In our case, this condition brings us to inequality for $p$, $q$ and $\lambda$, namely $\frac{q^{2}-r^{2}}{\lambda^{2}-1}-q^{2}>0$. This inequality does not work for $\lambda^2>2$. Thus, the eigenvalues of matrix $M$ are real for $\lambda^2<2$.        

Let now investigate our Hamiltonian in the same manner as the example above. For simplicity, we consider the case of $\boldsymbol{\vartheta}=(0,0,\vartheta)$. In this case, the Hamiltonian in Eq.~(10) from main text is written as 
\begin{equation}
	\label{Hamiltonian for vartheta directed along z}
	\mathcal{H}=\frac{\upsilon_{F}}{1-\beta^{2}}\left[\left(\begin{array}{cc}
		p_{z} & \left(1-\beta\right)\left(p_{x}-ip_{y}\right)\\
		\left(1+\beta\right)\left(p_{x}+ip_{y}\right) & -p_{z}
	\end{array}\right)+\beta p_{z} \sigma_0\right],
\end{equation} 
where $\beta=\vartheta/v_F$ and $\sigma_0$ is the identity matrix. The matrix $\eta$ has the form $\eta=\left(\begin{array}{cc}
	a & b\\
	b^{\ast} & d
\end{array}\right)$, where $a=\frac{r+bp_{z}}{\left(1-\beta\right)\left(p_{x}-ip_{y}\right)}$ and $d=\frac{r-bp_{z}}{\left(1+\beta\right)\left(p_{x}-ip_{y}\right)}$, $r,b\in\mathbb{C}$. From condition $\eta=S^{\dagger} S$ one gets that $q,r>0\in\mathbb{R}$ and the condition $\det(S^{\dagger}S)>0$ results in inequality for other parameters, $ad-\left|b\right|^{2}>0\in\mathbb{R}$. Using these conditions at $\beta<1$, one can show that matrix $\eta=S^{\dagger}S$ exists always. For example, $\eta=\left(\begin{array}{cc}
	\frac{q}{1-\beta} & 0\\
	0 & \frac{q}{1+\beta}
\end{array}\right)$, where $q>0\in\mathbb{R}$. For $\beta>1$ the matrix $\eta=S^{\dagger}S$ exists only for $\frac{p_{z}^{2}}{p_{x}^{2}+p_{y}^{2}}<\beta^{2}-1$, which turns out to be $|\sin \chi|<\beta^{-1}$ in spherical coordinates, and this condition coincides with the one we have provided in Eq.~(9) from main text. In this case the matrix has the form 
$ \eta=\left(\begin{array}{cc}
	\frac{p_{z}}{\beta-1} & -p_{x}+ip_{y}\\
	-p_{x}-ip_{y} & \frac{p_{z}}{\beta+1}
\end{array}\right)$. 
In other words, for $\left|\sin\chi\right|<\beta^{-1}$ the Hamiltonian above, Eq.~(\ref{Hamiltonian for vartheta directed along z}) can be brought to hermitian Hamiltonian using similarity transformations. Namely $S^{-1}\mathcal{H}S$ is the Hermitian operator, despite the fact that $\mathcal{H}$ is non-Hermitian. Additionally, it is worth pointing out, that redefinition of scalar product in Hilbert space (unitary theorem) in the following form $\left\langle \left\langle \Psi|\Psi\right\rangle \right\rangle _{S}\equiv \left\langle \Psi|S^{\dagger}S|\Psi\right\rangle$ does depend on time, if Hamiltonian satisfies the condition Eq.~(12) from main text. Thus, this means that the probability density does not depend on time, and the unitary condition is suited. 

As one can observe from Eq.~(12) from main text, in general the realness of spectrum is defined not only by the properties of operator by itself, but as well it influences on Hilbert space. It worth mentioning, that the $\mathcal{PT}$ symmetry for non-Hermitian quantum systems introduced by Bender~\cite{Bender}, says that $\mathcal{PT}$ operator and Hamiltonian have the same eigenvectors. In this case Hamiltonian has the real spectrum, despite the fact that it is non-Hermitian. This requirements is the particular case of Eq.~(12) from main text. Namely, to show this, let change a bit the matrix we have considered above, $M\Rightarrow\left(\begin{array}{cc}
	1 & \lambda-1\\
	\lambda+1 & -1
\end{array}\right)$. This matrix is not hermitian and it does not have $\mathcal{PT}$ symmetry. However, the eigenvalues of this matrix are real for any real $\lambda$. This is related with the fact that matrix $\left(\begin{array}{cc}
	1 & \lambda-1\\
	\lambda+1 & -1
\end{array}\right)$ is equivalent to hermitian matrix $\left(\begin{array}{cc}
	0 & \lambda\\
	\lambda & 0
\end{array}\right)$. Namely, there always exists the similarity transformation $R$, therefore $R\left(\begin{array}{cc}
	0 & \lambda\\
	\lambda & 0
\end{array}\right)R^{-1}=\left(\begin{array}{cc}
	1 & \lambda-1\\
	\lambda+1 & -1
\end{array}\right)$.

\section{D: Fractional Chern number}
The Chern number is given by 
\begin{equation}
	\label{Chern number. SM Section D}	
	\mathcal{N}_{\alpha}=\frac{1}{4\pi}\oint\bm{\Omega}_{\alpha}\cdot d\mathbf{S},
\end{equation}
where $\bm{\Omega}_{\alpha}$ is Berry curvature of band $\alpha$, and the integration is carried out with respect to closed surface, enclosing the Weyl point. We rewrite the Eq.~(\ref{Chern number. SM Section D}) as
\begin{equation}
	\mathcal{N}_{\alpha}=\frac{1}{4\pi}\left(\int_{S_{\textrm{C}}}\bm{\Omega}_{\alpha}\cdot d\mathbf{S}+\int_{S_{\textrm{R}}}\bm{\Omega}_{\alpha}\cdot d\mathbf{S}\right),
\end{equation}
where $S_{\textrm{C}}$ and $S_{\textrm{R}}$ are domains of surface corresponding to complex and real spectra (see Fig. 4 from main text). The Berry curvature for non-Hermitian Hamiltonian is given by 
\begin{equation}
	\bm{\Omega}_{\alpha}=i\partial_{\mathbf{p}}\times\left\langle \varPsi_{\alpha}^{L}\right|\partial_{\mathbf{p}}\left|\varPsi_{\alpha}^{R}\right\rangle.
\end{equation}
One can show, that $\int_{S_{\textrm{C}}}\bm{\Omega}_{\alpha}\cdot d\mathbf{S}=0$. To do this, we consider the time-evolution of eigenstates at $D=C$. In this case the spectrum has only one band, $\varepsilon_{R}=\varepsilon_{L}=A\left(\mathbf{p}\right)$ and the eigenstates are given by  
\begin{equation}
	\left|\varPsi_{\pm}^{R}\right\rangle =\left|\varPsi^{R}\right\rangle =\left(\begin{array}{c}
		\frac{C_{z}-iD_{z}}{C_{x}+i\left(C_{y}-D_{x}\right)+D_{y}}\\
		1
	\end{array}\right),\,\,\,\,\left|\varPsi_{\pm}^{L}\right\rangle =\left|\varPsi^{L}\right\rangle =\left(\begin{array}{c}
		\frac{C_{z}+iD_{z}}{C_{x}+i\left(C_{y}+D_{x}\right)-D_{y}}\\
		1
	\end{array}\right).
\end{equation}
From the above equations one gets $\left\langle \varPsi^{L}\right|\left|\varPsi^{R}\right\rangle =\left\langle \varPsi^{R}\right|\left|\varPsi^{L}\right\rangle =0$, and for adiabatic time evolution we have $i\partial_{t}\left|\varPsi^{R}\left(t\right)\right\rangle =\mathcal{H}\left(t\right)\left|\varPsi^{R}\left(t\right)\right\rangle =\varepsilon\left(t\right)\left|\varPsi^{R}\left(t\right)\right\rangle$. Therefore
\begin{equation}
	i\left\langle \varPsi^{L}\right|\partial_{t}\left|\varPsi^{R}\right\rangle =\varepsilon\left(t\right)\left\langle \varPsi^{L}\right|\left|\varPsi^{R}\right\rangle =0, \quad \text{or} \quad
	\left\langle \varPsi^{L}\right|\partial_{\mathbf{p}}\left|\varPsi^{R}\right\rangle \cdot\partial_{t}\mathbf{p}=0.
\end{equation}
Consequently, the line integral $\oint\left\langle \varPsi^{L}\right|\partial_{\mathbf{p}}\left|\varPsi^{R}\right\rangle \cdot d\mathbf{p}$, corresponding to equation $D=C$ vanishes. Indeed
\begin{equation}
	\oint\left\langle \varPsi^{L}\right|\partial_{\mathbf{p}}\left|\varPsi^{R}\right\rangle \cdot d\mathbf{p}=\int_{0}^{T}\left(\left\langle \varPsi^{L}\right|\partial_{\mathbf{p}}\left|\varPsi^{R}\right\rangle \cdot\partial_{t}\mathbf{p}\right)dt=0,
\end{equation}
where the cyclic evolution around closed path is denoted by times $0$ and $T$. The line (contour) for $D=C$ is presented at Fig.~4 in main text. According to Stoke's theorem
\begin{equation}
	\label{Contour integral SM Section D}	
	\oint_{\bm{l}_{1}+\bm{l}_{2}}\left\langle \varPsi_{\alpha}^{L}\right|\partial_{\mathbf{p}}\left|\varPsi_{\alpha}^{R}\right\rangle \cdot d\mathbf{p}=\int_{S_{\textrm{C}}}\left(\partial_{\mathbf{p}}\times\left\langle \varPsi_{\alpha}^{L}\right|\partial_{\mathbf{p}}\left|\varPsi_{\alpha}^{R}\right\rangle \right)\cdot d^{2}\mathbf{p}=0,
\end{equation}
since the contour $\bm{l}_{1}+\bm{l}_{2}$ is the  boundary of domain $S_{\textrm{C}}$ (see Fig.~4 in main text). It is worth mentioning, that the Stoke's theorem can be applied only to domian $S_{\textrm{C}}$, because the orthogonality condition $\left\langle \varPsi_{\alpha}^{L}|\varPsi_{\alpha}^{R}\right\rangle =0$ is the same as for contour $\bm{l}_{1}+\bm{l}_{2}$. Thus, the domain of momentum space, $S_{\textrm{C}}$, corresponding to complex spectrum does not contribute to integral, Eq.~(\ref{Contour integral SM Section D}). Consequently, it follows that the Chern number is given by
\begin{equation}
	\mathcal{N}_{\alpha}=\frac{1}{4\pi}\int_{S_{\textrm{R}}}\bm{\Omega}_{\alpha}\cdot d^{2}\mathbf{p}.
\end{equation}  
Thereby, only part of the sphere in momentum space contributes to integral, which results in fractional Chern number. Thus, the Weyl fermions of types III and IV are characterized by fractional Chern number.


\begin{thebibliography}{99}
	
\bibitem{Bansil}
Bansil, A. and Lin, Hsin and Das, Tanmoy, Rev. Mod. Phys. {\bf 88}, 021004 (2016)	

\bibitem{Gilbert}
Gilbert, Matthew J., Communications Physics {\bf 4}, 70 (2021)
	
\bibitem{Liu}
Z. K. Liu  and B. Zhou  and Y. Zhang  and Z. J. Wang  and H. M. Weng  and D. Prabhakaran  and S.-K. Mo  and Z. X. Shen  and Z. Fang  and X. Dai  and Z. Hussain  and Y. L. Chen, Science {\bf 343}, 864867 (2014)
	

\bibitem{Borisenko}
Borisenko, Sergey and Gibson, Quinn and Evtushinsky, Danil and Zabolotnyy, Volodymyr and B\"uchner, Bernd and Cava, Robert J., Phys. Rev. Lett. {\bf 113}, 027603 (2014)


\bibitem{Yang}
Yang, Bohm-Jung and Nagaosa, Naoto, Nature Communications {\bf 5}, 4898 (2014) 


\bibitem{Xu}
Su-Yang Xu  and Chang Liu  and Satya K. Kushwaha  and Raman Sankar  and Jason W. Krizan  and Ilya Belopolski  and Madhab Neupane  and Guang Bian  and Nasser Alidoust  and Tay-Rong Chang  and Horng-Tay Jeng  and Cheng-Yi Huang  and Wei-Feng Tsai  and Hsin Lin  and Pavel P. Shibayev  and Fang-Cheng Chou  and Robert J. Cava  and M. Zahid Hasan, Science {\bf 347}, 294298 (2015)
	

\bibitem{Vafek}
Vafek, Oskar and Vishwanath, Ashvin, Annual Review of Condensed Matter Physics {\bf 5}, 83112 (2014)


\bibitem{Weng}
Weng, Hongming and Fang, Chen and Fang, Zhong and Bernevig, B. Andrei and Dai, Xi, Phys. Rev. X {\bf 5}, 011029 (2015)


\bibitem{Huang}
Huang, Shin-Ming and Xu, Su-Yang and Belopolski, Ilya and Lee, Chi-Cheng and Chang, Guoqing and Wang, BaoKai
		and Alidoust, Nasser
		and Bian, Guang
		and Neupane, Madhab
		and Zhang, Chenglong
		and Jia, Shuang
		and Bansil, Arun
		and Lin, Hsin
		and Hasan, M. Zahid, Nature Communications {\bf 6}, 7373 (2015)

\bibitem{Nasser}
Su-Yang Xu  and Ilya Belopolski  and Nasser Alidoust  and Madhab Neupane  and Guang Bian  and Chenglong Zhang  and Raman Sankar  and Guoqing Chang  and Zhujun Yuan  and Chi-Cheng Lee  and Shin-Ming Huang  and Hao Zheng  and Jie Ma  and Daniel S. Sanchez  and BaoKai Wang  and Arun Bansil  and Fangcheng Chou  and Pavel P. Shibayev  and Hsin Lin  and Shuang Jia  and M. Zahid Hasan, Science {\bf 349}, 613617 (2015)


\bibitem{Lv}
Lv, B. Q. and Weng, H. M. and Fu, B. B. and Wang, X. P. and Miao, H. and Ma, J. and Richard, P. and Huang, X. C. and Zhao, L. X. and Chen, G. F. and Fang, Z. and Dai, X. and Qian, T. and Ding, H., Phys. Rev. X {\bf 5}, 031013 (2015)


\bibitem{Bradlyn}
Barry Bradlyn  and Jennifer Cano  and Zhijun Wang  and M. G. Vergniory  and C. Felser  and R. J. Cava  and B. Andrei Bernevig, Science {\bf 353} aaf5037 (2016)


\bibitem{Bzdusek}
Bzdu{\v{s}}ek, Tom{\'a}{\v{s}} and Wu, QuanSheng
and R{\"u}egg, Andreas
		and Sigrist, Manfred
		and Soluyanov, Alexey A., Nature {\bf 538} 7578 (2016)
	


\bibitem{Rui}
Yu, Rui and Weng, Hongming and Fang, Zhong and Dai, Xi and Hu, Xiao, Phys. Rev. Lett. {\bf 115}, 036807 (2015)


\bibitem{Bian}
Bian, Guang
		and Chang, Tay-Rong
		and Sankar, Raman
		and Xu, Su-Yang
		and Zheng, Hao
		and Neupert, Titus
		and Chiu, Ching-Kai
		and Huang, Shin-Ming
		and Chang, Guoqing
		and Belopolski, Ilya
		and Sanchez, Daniel S.
		and Neupane, Madhab
		and Alidoust, Nasser
		and Liu, Chang
		and Wang, BaoKai
		and Lee, Chi-Cheng
		and Jeng, Horng-Tay
		and Zhang, Chenglong
		and Yuan, Zhujun
		and Jia, Shuang
		and Bansil, Arun
		and Chou, Fangcheng
		and Lin, Hsin
		and Hasan, M. Zahid, 
Nature Communications {\bf 7} 10556 (2016)


\bibitem{VolovikBook}
Volovik, G. E. \textit{The Universe in a Helium Droplet}, (Claverdon Press, Oxford, 2003).


\bibitem{Herring}
Herring, Conyers, Phys. Rev. {\bf 52} 365373 (1937)


\bibitem{Armitage}
Armitage, N. P. and Mele, E. J. and Vishwanath, Ashvin, Rev. Mod. Phys. {\bf 90}, 015001 (2018)

\bibitem{Zubkov}
G.E. Volovik and M.A. Zubkov, Nuclear Physics B {\bf 881}, 514538 (2014)

\bibitem{Soluyanov}
Soluyanov, Alexey A.
		and Gresch, Dominik
		and Wang, Zhijun
		and Wu, QuanSheng
		and Troyer, Matthias
		and Dai, Xi
		and Bernevig, B. Andrei,
Nature {\bf 527}, 595498 (2015)


\bibitem{Sun}
Sun, Yan and Wu, Shu-Chun and Ali, Mazhar N. and Felser, Claudia and Yan, Binghai, 
Phys. Rev. B {\bf 92}, 161107 (2015)

\bibitem{Wang}
Wang, Zhijun and Gresch, Dominik and Soluyanov, Alexey A. and Xie, Weiwei and Kushwaha, S. and Dai, Xi and Troyer, Matthias and Cava, Robert J. and Bernevig, B. Andrei, Phys. Rev. Lett. {\bf 117}, 056805 (2016)

\bibitem{Belopolski}
Belopolski, Ilya
		and Sanchez, Daniel S.
		and Ishida, Yukiaki
		and Pan, Xingchen
		and Yu, Peng
		and Xu, Su-Yang
		and Chang, Guoqing
		and Chang, Tay-Rong
		and Zheng, Hao
		and Alidoust, Nasser
		and Bian, Guang
		and Neupane, Madhab
		and Huang, Shin-Ming
		and Lee, Chi-Cheng
		and Song, You
		and Bu, Haijun
		and Wang, Guanghou
		and Li, Shisheng
		and Eda, Goki
		and Jeng, Horng-Tay
		and Kondo, Takeshi
		and Lin, Hsin
		and Liu, Zheng
		and Song, Fengqi
		and Shin, Shik
		and Hasan, M. Zahid,
Nature Communications {\bf 7}, 13643 (2016)


\bibitem{Kedem}
Kedem, Yaron and Bergholtz, Emil J. and Wilczek, Frank, 
Phys. Rev. Research {\bf 2}, 043285 (2020)

\bibitem{Thomas}
Christophe De Beule and Solofo Groenendijk and Tobias Meng and Thomas L. Schmidt, arXiv:2106.14595 (2021) 

\bibitem{Nissinen}
Nissinen, J. and Volovik, G. E., JETP Lett. {\bf 105} 442 (2017)

\bibitem{Nissinen2}
Nissinen, J. and Volovik, G. E., Journal of Experimental and Theoretical Physics {\bf 127} 948957 (2018)

\bibitem{Brody}
Dorje C Brody, Journal of Physics A: Mathematical and Theoretical {\bf 47}, 035305 (2013)

\bibitem{Emil}
Bergholtz, Emil J. and Budich, Jan Carl and Kunst, Flore K.,
Rev. Mod. Phys. {\bf 93}, 015005 (2021)

\bibitem{Ghatak}
Ananya Ghatak and Tanmoy Das, J. Phys.: Condens. Matter {\bf 31}, 263001 (2019)

\bibitem{SM}
Supplemental Material at [] for details of calculations of eigenstates, scalar products and fractional Chen number

\bibitem{Ali1}
Mostafazadeh, Ali, Journal of Mathematical Physics {\bf 43}, 205214 (2002)

\bibitem{Ali2}
Mostafazadeh, Ali, Journal of Mathematical Physics {\bf 43}, 28142816 (2002)

\bibitem{Ali3}
Mostafazadeh, Ali, Journal of Mathematical Physics {\bf 43}, 39443951 (2002)

\bibitem{Fu}
Shen, Huitao and Zhen, Bo and Fu, Liang, Phys. Rev. Lett. {\bf 120}, 146402 (2018)

\bibitem{TonyLee}
Lee, Tony E., Phys. Rev. Lett. {\bf 116}, 133903 (2016)

\bibitem{Landsberg}
E. Antonc{\'{\i}}k and P. T. Landsberg, Proceedings of the Physical Society {\bf 82}, 337342 (1963)


\bibitem{Halpern}
V. Halpern, Journal of Physics and Chemistry of Solids {\bf 24}, 14951502 (1963)

\bibitem{Bernstein}
Bernstein, N. and Mehl, M. J. and Papaconstantopoulos, D. A., Phys. Rev. B {\bf 66}, 075212 (2002)

\bibitem{Boykin}
Boykin,Timothy B.  and Sarangapani,Prasad  and Klimeck,Gerhard, Journal of Applied Physics {\bf 125}, 144302 (2019)

\bibitem{Benjamin}
Thomas Benjamin Smith and Alessandro Principi, Physica E: Low-dimensional Systems and Nanostructures {\bf 126}, 114423 (2021)	
	
\end{thebibliography}

\begin{thebibliography}{99}
	
	\bibitem{Bender}
	C. M. Bender, \textit{PT Symmetry In Quantum and Classical Physics,} (World Scientific, Singapore, 2019)
	
\end{thebibliography}
\end{document}